\documentclass[showpacs,preprintnumbers,amsmath,amssymb, 
eqsecnum,
twocolumn, tightenlines,
]{revtex4}

\usepackage{bm}

\usepackage{amsmath}

\usepackage{dcolumn}

   \usepackage[dvips]{graphicx}
\begin{document}

    \bibliographystyle{apsrev}
    
    \title {Reflection on event horizon and escape of particles from
      confinement inside black holes}
    
    \author{M.Yu.Kuchiev}
    \email[Email:]{kuchiev@newt.phys.unsw.edu.au}
    \author{V.V.Flambaum}
    \email[Email:]{flambaum@newt.phys.unsw.edu.au}
    
    \affiliation{School of Physics, University of New South Wales,
      Sydney 2052, Australia}
    
    \date{\today}

    \begin{abstract}
      Several recently found properties of the event horizon of black
      holes are discussed. One of them is the reflection of
      the incoming particles on the horizon. A particle approaching
      the black hole can bounce on the horizon back, into the outside
      world, which drastically reduces the absorption cross section in
      the infrared region.  Another, though related phenomenon takes
      place for particles inside the horizon. A locked inside particle
      has, in fact, an opportunity to escape into the outside world.
      Thus, the confinement inside the horizon is not absolute.  The
      escape from within the interior region of the horizon allows the transfer
      of information from this region into the outside world.
      This result may help resolve the information paradox for
      black holes. Both the reflection and escape phenomena happen due
      to pure quantum reasons, being impossible in the classical
      approximation.
    \end{abstract}
    
    \pacs{04.70.Dy, 04.20.Gz}

    \maketitle
    \section{introduction}    
    \label{intro}
    
    A progress related to several quantum phenomena that take place on
    the event horizon of black holes is outlined. One of the effects
    discussed here is the well known Hawking radiation.  The other two
    phenomena examined were found only recently. They are the
    reflection on the event horizon of black holes, and the escape of
    particles from the confinement within the inside region of the
    horizon into the outside world.
    
    The effect of the Hawking radiation \cite{hawking_74,hawking_75},
    and the closely related Unruh process \cite{unruh}, have links
    with the entropy of black holes studied by Bekenstein
    \cite{bekenstein_1972,bekenstein_1973,bekenstein_1974}.  For a
    recent review on the Hawking effect and thermodynamics properties
    of black holes see \cite{jakobson_2003,wald_2001}.  Treatment of
    quantum phenomena on the event horizon, including the problem of
    the origin of the entropy of black holes, the density of its
    quantum states, and the brick wall model can be found in the
    reviews by 't Hooft \cite{thooft_01,thooft_96}, see also his Ref.
    \cite{thooft_85}.
     
    The phenomenon of reflection on the event horizon, which will be
    discussed in detail in this work, has strong connections with the
    scattering problem.  The first analytical results in the
    scattering problem were obtained by Starobinsky
    \cite{starobinsky_73} for the scalar field and Starobinsky and
    Churilov \cite{starobinsky_churilov_73} for electromagnetic and
    gravitational waves scattered by the rotating Kerr black hole.
    Independently, Unruh \cite{unruh_76} considered scattering of
    scalar and fermion particles by Schwarzschild black holes.  A
    detailed study of the scattering problem was given by Sanchez
    \cite{sanchez_1977,sanchez_1997}.  After these and a number of
    subsequent works, it has been assumed that the scattering problem
    is completely understood, see details and bibliography in
    the books \cite{frolov_novikov_98,thorne_1994,chandrasekhar_93,%
      fullerman_handler_matzner_88}.
            
    However, it was found recently in Refs.
    \cite{kuchiev_1,kuchiev_2,kuchiev_3} that a particle approaching
    the black hole can bounce on the horizon back into the outside
    world. This phenomenon is referred to below as the reflection on the
    horizon (RH).  The effect arises due to pure quantum reasons,
    being obviously absent in the classical approximation.  The RH
    prompts a strong decrease in the absorption cross section,
    reducing it to a zero value in the infrared region, as was shown
    explicitly in \cite{kuchiev_flambaum_04} for scattering of scalar
    massless particles on Schwarzschild black holes.  This behavior
    of the cross section differs qualitatively from the previously
    accepted result, which stated that the low energy absorption cross
    section equals the area of the horizon \cite{unruh_76}.  The
    similar reduction of the cross section is expected to take place
    for scattering of any massless particle by any black hole in the
    low energy limit.
    
    The quantum phenomena that prompt the existence of the RH have
    another unexpected manifestation, for particles confined inside
    the horizon.  Classically this confinement is absolute.  However,
    the quantum treatment of the problem in Refs.
    \cite{kuchiev_1,kuchiev_2,kuchiev_3} revealed that the wave
    function of any confined particle necessarily has a particular
    admixture that behaves on the horizon as the outgoing wave. This
    property of the wave function indicates that any confined particle
    has a chance to escape from the inside region into the outside
    world.  Thus, the confinement inside a black hole is not perfect,
    a particle locked inside has a chance to find its way out. We will
    refer to this opportunity as the escape effect (EE).  The Hawking
    radiation process can be considered as a manifestation of the EE
    in particular circumstances, when a black hole is put inside the
    temperature bath.  However, in the general case the EE allows the
    extraction of information from the inside region of the horizon
    into the outside world.
    
    This work briefly summarizes some of the arguments of
    \cite{kuchiev_1,kuchiev_2,kuchiev_3,kuchiev_flambaum_04} related
    to the RH and EE. Units $\hbar =c=2GM=1$ are used, where $G$ and
    $M$ are the gravitational constant and the black hole mass; the
    gravitational radius in these units reads $r_g = 2GM/c^2=1$.
    
    \section{Reflection on horizon}
    \label{reflection}
    
    Let us formulate briefly the results of
    Refs.\cite{kuchiev_1,kuchiev_2,kuchiev_3}. Consider for simplicity
    the scalar massless field in the vicinity of the Schwarzschild
    black hole.  Take a scalar particle with the energy $\varepsilon$
    and zero orbital momentum $l=0$. It is easy to verify (see e. g.
    \cite{frolov_novikov_98,thorne_1994,chandrasekhar_93,fullerman_handler_matzner_88},
    or Eqs.(\ref{in}),(\ref{outout}) below)
    that the corresponding wave function exhibits behavior $\phi (r)
    \simeq \exp[\,\mp i \varepsilon \,\ln\,(r-1)\,]$ on the horizon
    $r\rightarrow 1$, where the signs minus and plus describe the
    waves propagating inside and outside of the horizon respectively.
    The general form of the wave function on the horizon is therefore
    \begin{eqnarray}
      \label{gen}
      \phi(r)\simeq 
    \exp[- i \varepsilon \ln (r-1)\,] 
    + \mathcal{R}
    \exp[ i \varepsilon \ln (r-1)\,]~. 
    \end{eqnarray}
    Suppose we consider the scattering problem, the impact of scalar
    particles on the black hole. Then the first term in Eq.(\ref{gen})
    is definitely present, it describes the flux of incoming
    particles. It seems also {\it natural} to expect that there is no
    second term, because the horizon is {\it presumed} to be a prefect
    absorber. In other words, it seems natural to put in
    Eq.(\ref{gen}) $\mathcal{R}=0$.  Exactly this condition has always
    been used in the scattering problem for different particles
    (scalars, spinors, electromagnetic and gravitational waves) and
    different types of black holes (Schwarzschild, Kerr and others),
    see the pioneering Refs.
    \cite{starobinsky_73,starobinsky_churilov_73,unruh_76}, later
    developments can be found in the review \cite{sanchez_1997} and
    books
    \cite{frolov_novikov_98,chandrasekhar_93,fullerman_handler_matzner_88}.
        
    It was unexpected therefore that Refs.
    \cite{kuchiev_1,kuchiev_2,kuchiev_3} argued that $\mathcal{R}$ in
    Eq.(\ref{gen}) has, in fact, a nonzero value, specifically that
    \begin{eqnarray}
      \label{R}
      | \mathcal{R} | = \exp\left(-\frac{\varepsilon}{2T}\,\right)~,
    \end{eqnarray}
    where $T=1/(4\pi)$ is the Hawking temperature. The fact that
    $|\mathcal{R}|>0$ means that there is a reflected wave in
    Eq.(\ref{gen}) that gives rise to the flux of outgoing particles.
    Correspondingly, $\mathcal{R}$ is to be called the reflection
    coefficient.  The fact that it is nonzero indicates that a
    particle can bounce on the horizon back into the outside world.
    Eq.(\ref{R}) presents an explicit form for the effect that was
    called the RH in Section \ref{intro}. For low energies
    $\varepsilon < T$ the RH is very effective, which makes the
    horizon a good {\it reflector} in the infrared region.  This
    property is in contrast with the conventional point of view that
    presumes the horizon to be a perfect absorber.
    
    Eq.(\ref{R}) has a profound influence on the absorption cross
    section for low energies of incoming particles $\varepsilon \le
    T$. Assuming conventional properties of the horizon ($\mathcal{R}
    =0$) Unruh found \cite{unruh_76} that in the infrared region the
    cross section equals the area of the horizon
    \begin{eqnarray}
      \label{Un}
      \sigma_\mathrm{abs} = 4 \pi r_g^2~, \quad \varepsilon
      \rightarrow 0~.
    \end{eqnarray}
    (Here and in Eq.(\ref{kf}) below conventional units are used.)
    Taking the RH into account Ref.
    \cite{kuchiev_flambaum_04} arrived at a different result
    \begin{eqnarray}
      \label{kf}
      \sigma_\mathrm{abs} = 4 \pi^2 \varepsilon r_g^3/(\hbar c)~, \quad \varepsilon
      \rightarrow 0~,
    \end{eqnarray}
    which means that the cross section vanishes for low energy. 
    
    There is an appealing physical picture suggested in
    \cite{kuchiev_flambaum_04} that describes the RH as a creation of a
    pair at the horizon followed by an annihilation of one of the created
    particles with the inner particle inside the black hole.  This is
    close to the usual physical explanation of the Hawking effect via
    the pair production \footnote{Note, however, that the simple
      mechanism of the pair production possesses a difficulty, as is
      discussed in some detail in Section \ref{inform} below.  At this
      point, however, we stick to the conventional point of view,
      neglecting this complication in order to present the RH in the
      commonly used terms.}.

    Eqs.(\ref{gen}),(\ref{R}) and (\ref{kf}) summarize the main claims
    that Refs.
    \cite{kuchiev_1,kuchiev_2,kuchiev_3,kuchiev_flambaum_04} made for
    the outside region. Later on, in Section \ref{inside}, we will
    discuss the implications of these results for the inside region.

    \section{Large wavelengths}
    \label{gribov}
  
    The book of Khriplovich \cite{khriplovich} mentions qualitative
    arguments put forward by Gribov in early 70's, which indicated
    that black holes are capable of radiating. One of his reasons, as
    the book presents it, was that ``it is obvious that a black hole
    is incapable of containing radiation with the wavelength exceeding
    the gravitational radius'' (p. 112 of \cite{khriplovich}, in our
    translation from Russian).  This argument may look simplistic
    (though it was not the only one articulated by Gribov), but
    keeping in mind that it was made before the Hawking finding, its
    simplicity bears, arguably, an aura of a classical foreseeing.
    
    If one allows oneself to rely on this argument in the scattering
    problem, one has to conclude that the absorption of particles with
    large wavelengths by a black hole should meet a difficulty, in
    other words it should be suppressed. This is exactly what
    Eq.(\ref{kf}) which takes the RH into account predicts. In
    contrast, Eq.(\ref{Un}) which neglects the RH shows no sign of
    such suppression.
    
    Thus, one may argue that the RH is in line with the Gribov
    argument related to large wavelengths.

    \section{Reflection on horizon as above-barrier reflection} 
    \label{AT-reflection}
    
    Let us discuss a simple argument used in Refs.
    \cite{kuchiev_2,kuchiev_flambaum_04} to justify validity of the
    RH.  Consider the Schwarzschild geometry with the metric
    \begin{eqnarray}
      \label{schw}
      ds^2 = -\left(1-\frac{1}{r}\right)dt^2+\frac{dr^2}{1-1/r}+r^2 d\Omega^2~,     
    \end{eqnarray}
    where $d\Omega^2$ describes the contribution of angular variables.
    Take the scalar field, assuming for simplicity that it is
    massless. Choose the most important for us wave with the zero
    orbital momentum $l=0$. Then the radial wave function $\phi(r)$
    for the stationary state with energy $\varepsilon$ satisfies the
    Klein-Gordon equation
      \begin{eqnarray}
        \label{phi''}
        \phi''(r) + \left(\frac{1}{r}+\frac{1}{r-1} \right) \phi'(r)
        + \frac{\varepsilon ^2}{(1-1/r)^2}  \phi(r) = 0.
      \end{eqnarray}
      Making the substitution $\phi(r) \rightarrow \psi(r)=
      [\,r(r-1)\,]^{1/2} \phi(r)$ one can rewrite Eq.(\ref{phi''})
      \begin{eqnarray}
        \label{p2}
        \varepsilon^2 \psi(r) = - \psi''(r)+U(r)\psi(r)~,
      \end{eqnarray}
      where
      \begin{eqnarray}
        \label{UU}
        U(r) = -\frac{1}{ (r-1)^2}\left(\varepsilon^2 + \frac{1}{4r^2}\right)
        - \frac{2\varepsilon^2}{r-1},
      \end{eqnarray}
      reducing it to the form of the conventional Schr\"odinger-type
      equation, if $U(r)$ is considered as an effective,
      energy-dependent potential (note that it is strictly attractive),
      and $\varepsilon^2$ on the left-hand side is accepted as the
      eigenvalue.
      
      Consider the scattering problem, the impact of scalar particles
      on the black hole. Then, definitely there is the incoming wave
      that falls on the horizon. In the proximity of the horizon
      $|r-1| \ll 1$ this wave reads
      \begin{eqnarray}
        \label{in}
        \phi_\mathrm{in}(r) = \exp[-i\varepsilon \ln(r-1)\,],\quad
        r>1~,
      \end{eqnarray}
      as can be verified using Eq.(\ref{phi''}), or (\ref{p2}).  
      
      Let us look at the problem from the traditional point of view.
      There is the incoming wave Eq.(\ref{in}), and there is the potential
      Eq.(\ref{UU}).  One can expect therefore that there should exist
      also the outgoing wave, which is always present in quantum
      mechanical problems of this type.  This is true even for
      attractive potentials. The only distinction for the attractive
      potentials is that the reflection for them is prompted by pure
      quantum reasons, being absent in the classical approximation.  As
      a result, the reflection coefficient is to be exponentially
      small.  In the problem at hand, the potential $U(r)$ in
      Eq.(\ref{UU}) is smooth in the region $r>1$, the semiclassical
      approximation works well for it, with the only exception of the
      horizon $r=1$, where the potential has a singular point
      $U(r)\simeq - (\varepsilon^2+1/4)/(r-1)^2$.
      
      We can apply therefore conventional semiclassical methods by
      taking the incoming wave Eq.(\ref{in}) and continuing it into the
      region $r<1$ by means of the analytical continuation over the
      lower semiplane of the complex plane $r$ that avoids the
      singularity at $r=1$.  The result, which reads
      \begin{eqnarray}
        \label{inin}
        \phi_\mathrm{in}(r) 
        = \exp[-\pi \varepsilon -i\varepsilon \ln(1-r)\,],\quad r<1~,
      \end{eqnarray}
      shows that the incoming wave exists in the inside region $r<1$.
      We need to determine therefore what happens to the wave function
      at the origin $r=0$, where, generally speaking, it behaves as
      $a+b\ln r$.  As usual, a solution regular at the origin should
      be chosen.  Such a solution cannot be constructed from the
      incoming wave Eq.(\ref{inin}), which necessarily incorporates
      the part singular at $r=0$.  We derive from this fact that in
      the region $r<1$ there should exist also the outgoing wave,
      which combines with the incoming wave to make the total wave
      function regular at the origin. In the vicinity of the horizon
      the outgoing wave can be presented as
      \begin{eqnarray}
        \label{outin}
        \phi_\mathrm{out}(r) 
        = \exp[-\pi \varepsilon +i\varepsilon \ln(1-r)+i\alpha\,]~,\quad r<1.
      \end{eqnarray}
      It has the same magnitude as the incoming wave (to allow a
      compensation of their singular parts at the origin), shifted, possibly,
      by a phase $\alpha$ that depends on details
      of the wave propagation far away from the horizon.
      
      This outgoing wave can now be continued into the region $r>1$ by
      using (again) the analytical continuation over the lower
      semiplane of the complex plane $r$.  As a result we find that
      there exists the outgoing wave in the outside region
      \begin{eqnarray}
        \label{outout}
        \phi_\mathrm{out}(r) 
        = \mathcal{R} \exp[\,i\varepsilon \ln(r-1)\,]~,\quad r>1,
      \end{eqnarray}
      where the coefficient is 
      \begin{eqnarray}
        \label{RR}
        \mathcal
      {R}=\exp(-2\pi\varepsilon+i\alpha)~.  
    \end{eqnarray}
    We come to the important conclusion. Alongside the incoming
    wave Eq.(\ref{in}), the wave function necessarily incorporates
    also the outgoing wave Eq.(\ref{outout}), in agreement with
    Eq.(\ref{gen}). The value for the reflection coefficient Eq.(\ref{RR})
    found here supports Eq.(\ref{R}).  This reflection
    coefficient is exponentially small for high energies, in accord
    with the naive anticipation for scattering on an
    attractive potential \footnote{In deriving this result, the
      continuation over the lower semiplane of the complex plane $r$
      was used.  If one applies a different analytical continuation,
      over the upper semiplane, one ends up with the large reflection
      coefficient $|\mathcal{R}|\ge 1$ that is physically
      unacceptable.  This fact can be used as an indication that the
      chosen way for the analytical continuation is correct.  Detailed
      validation of analytical properties of the wave function should
      be based on the causality condition, but we do not go into these
      details here, see more on the subject in
      \cite{kuchiev_1,kuchiev_2,kuchiev_3}}.
    
    In conclusion, the effective attractive potential Eq.(\ref{UU}),
    which is associated with the horizon, is able to reflect the
    incoming wave, which means that the RH really takes place. The
    effect has a similarity with the well known quantum phenomenon of
    the above-barrier reflection. The methods used here for its
    derivation are close to the conventional semiclassical treatment
    of the above-barrier reflection.

    \section{discrete symmetry of Schwarzschild geometry}
    
    Let us discuss another argument in favor of the RH, which was
    presented in \cite{kuchiev_1,kuchiev_2}. Consider the wave
    function $\phi(r)$ as an analytical function defined on the
    complex plane $r$. Take the real, physical value for $r$ in the
    vicinity of the horizon, $r>1,~r-1\ll 1$, where Eq.(\ref{gen}) is
    valid; then rotate $r$ around the point $r=1$ on the complex plane
    $r$ over an angle of $2\pi$ clockwise \footnote{This procedure
      resembles partially the method discussed in Section
      \ref{reflection}, where we applied similar rotation on the
      complex plane.  However in Section \ref{reflection} the angle of
      rotation was chosen $\pi$, which brought $r$ into the inside
      region $r<1$.  In contrast, the $2\pi$ rotation discussed here
      ends up with $r$ returning into the outside region $r>1$.}.
    Since we can keep $|r-1| \ll 1$, we can rely on Eq.(\ref{gen})
    throughout this transformation.  The transformation results in a
    new wave function $\tilde \phi(r)$
    \begin{eqnarray}
      \label{nv}
      \tilde \phi(r) =\varrho\, 
    \exp[- i \varepsilon \ln (r-1)]
    + \frac{\mathcal{R}}{\varrho}\,
    \exp[i \varepsilon \ln (r-1)],
    \end{eqnarray}
    where
    \begin{eqnarray}
      \label{rho}
      \varrho = \exp \left(-2\pi\,\varepsilon\right)~.
    \end{eqnarray}
    The important feature of the problem is the discrete symmetry of
    the Schwarzschild geometry. It can be expressed as a condition on
    functions $\phi(r)$ and $\tilde \phi(r)$
    \begin{eqnarray}
      \label{sym}
      [\,\tilde \phi(r)\,]^* = \exp(-i\alpha) \phi(r)~,
    \end{eqnarray}
    where $\alpha$ is a phase, which is not determined by this
    condition.  The origin and physical meaning of this symmetry are
    discussed below, see Eq.(\ref{AA}).  Meanwhile, to conclude the
    argument, note that from Eqs.(\ref{sym}),(\ref{nv}) one
    immediately finds that the reflection coefficient $\mathcal{R}$
    satisfies Eqs.(\ref{RR}) and (\ref{R}), thus verifying the RH.
    
\begin{figure}[tbh]
  \centering
  \includegraphics[height=8cm,keepaspectratio=true]{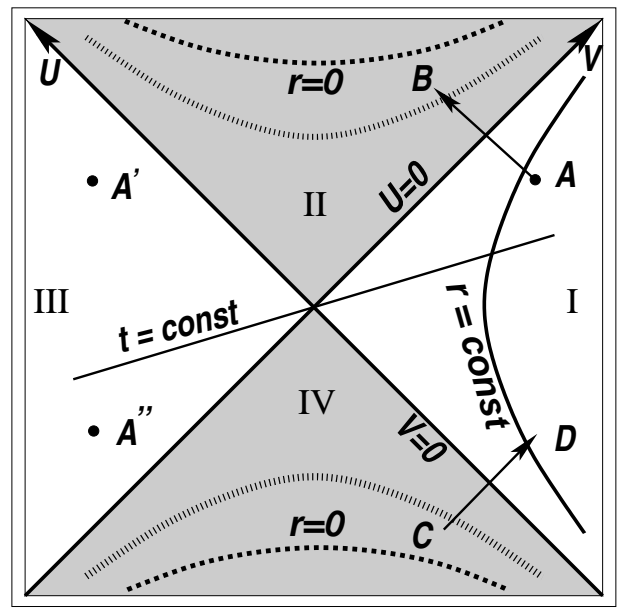}
\caption{\it 
  Kruskal coordinates. Areas I and III represent two identical copies
  of the outside region; II, IV show two inside regions. Hyperbolic
  curves $UV = const$ describe condition $r=const$, the dotted curve
  shows location of $r=0$, the inclined straight line presents
  condition $t=const$. The direction of time flow in I and III is
  opposite. The incoming particle follows {\it AB} crossing the
  horizon $U=0$ and residing in II. The outgoing particle {\it CD}
  escapes from IV crossing the horizon $V=0$ and coming to I.  Areas
  II and IV are not connected, which ensures classical confinement in
  II.  The wave function (\ref{gen}) (or (\ref{inout})) describe
  mixing of events that correspond to incoming and outgoing classical
  trajectories ({\it AB} and {\it CD}), resulting in phenomena of the
  reflection on horizon (RR) and the escape from the inside region
  (EE).  The symmetrically located points {\it A,A',A''} are used to
  reveal the symmetry (\ref{sym}) of the space-time.  The wave
  function (\ref{gen}) describes mixing of events corresponding to
  incoming and outgoing trajectories ({\it AB\/} and {\it CD\/}),
  which results in phenomena of RH and EE.  }
\label{one}
    \end{figure}

    It is convenient to look at the presented argument using Kruskal
    \cite{kruskal_1960} coordinates $U,V$
    \begin{eqnarray}
      \label{U}
      U&=&-(r-1)^{1/2} \,\exp \left( \frac{r-t}{2} \right)~,
      \\ \label{V}
      V&=&\,\,\,\,(r-1)^{1/2} \,\exp \left( \frac{r+t}{2} \right)~.
    \end{eqnarray}
    They are shown in Fig. \ref{one} in the conventional form, see e.g.
    Ref.\cite{misner_thorne_wheeler_1973}.  One observes that the
    rotation over the angle $2\pi$ around the point
    $r=1$ on the complex plane $r$ described above leads to the transformation
    $U\rightarrow U''=-U,~V\rightarrow V''=-V$, which brings the point
    $A$ on the Kruskal plane in Fig. \ref{one} to the point $A''$
    in the region III via the complex intermediate values of the
    variables $U,V$.
    The regions I ($U<0,~V>0$) and III ($U>0,~V<0$) of the Kruskal
    plane describe events that take place in the outside world
    \cite{misner_thorne_wheeler_1973}. These two regions are known to
    be identical, which provides an opportunity to describe each event
    in the outside world by one of the two points, either by the one
    located in the region I, or by the other one located in the region
    III.
    
    If some event is described by a point $A$ that has the coordinates
    $U,V$ in the region I, then the point $A'$, which represents the
    same event in the region III, has the coordinates $U',V'$, where
    $U'=V,~V'=U$, see Fig. \ref{one}. This identification of the
    points $A$ and $A'$ is particularly transparent in the vicinity
    of the event horizon that surrounds the internal region II in
    which all classical trajectories lead towards the singularity at
    $r=0$. If a particle follows the incoming trajectory in the region
    I, then it crosses the horizon $U=0,~V>0$, which separates the
    region I from region II. In this case the point $A$ accounts for
    an event that happens just before the particle reaches the
    horizon.  Alternatively, one can describe the incoming classical
    trajectory as the one that leads from the region III to region II.
    In that case the point $A'$ shows the event that precipitates
    the crossing of the horizon, which is located at $V=0,~U>0$. This
    description is in line with the fact that the motion on the
    Kruskal plane takes place ``from bottom to top''
    \footnote{Generally speaking, other identifications between the
      regions I and III are possible, see Refs.
      \cite{gibbons_86,chamblin_gibbons_95}.  However, the
      identification discussed here is most common, it is in
      line with the book \cite{misner_thorne_wheeler_1973}.}.

    In order to describe this motion one should choose appropriately a
    variable for the physical time.  In the region I the time can be
    described conventionally, with the help of the variable $t$. In
    contrast, the physical time in the region III should be described
    by the variable $\tilde t$, where $\tilde t=-t$.
    
    The described properties of the Kruskal plane have an important
    consequence. Since two points $A$ and $A'$ on the Kruskal plane
    describe one and the same event in the physical world, the wave
    function in these two points must have the same, up to a phase
    factor, value
    \begin{eqnarray}
      \label{AA}
      \phi(A) = \exp(-i\alpha)\phi(A')~.
    \end{eqnarray}
    where notation $\phi(A) \equiv \phi(r,t)$ is used, and the phase
    $\alpha$ does not depend on $A$.  Equation(\ref{AA}) represents a
    simple, but important symmetry condition related to the quantum
    properties of propagation in the Schwarzschild geometry. Its
    origin can be traced to the fact that the full coordinate system
    necessarily double-covers the Schwarzschild geometry.  The Kruskal
    coordinates provide a simple way to implement this fundamental
    property, though in deriving the symmetry condition Eq.(\ref{AA}) one
    may rely on any other full system of coordinates.
    
    Let us return now to a set of three points $A,A',A''$ shown on the
    Kruskal plane in Fig. \ref{one}; among them $A$ is the initial
    point, while $A''$ appears after the complex transformation of the
    variable $r$. This transformation leaves the time variable $t$
    intact, therefore at the point $A''$ we have $t''=t$.  Having in
    mind the symmetry condition Eq.(\ref{AA}), it is desirable to
    transform the point $A''$ into $A'$. This transformation amounts
    simply to the inversion of time because at the point $A'$ we have
    $t'=-t''$ (this inversion of time is in line with the property of
    the physical time $\tilde t =-t$ discussed above). The time
    inversion presumes the complex conjugation of the wave function.
    
    Returning now to Eq.(\ref{sym}) one observes that its
    left-hand side includes the wave function $\tilde \phi(r)$, which
    can be associated with the event that takes place at the point
    $A''$.  The complex conjugation of this function in Eq.(\ref{sym})
    gives another wave function, the one that is associated with
    the event that happens at the point $A'$ on the Kruskal plane.
    The symmetry condition Eq.(\ref{AA}) validates the identity
    between this later wave function and the initial wave function,
    which describes the event at the point $A$.
    
    In conclusion, the discrete symmetry of the Schwarzschild geometry
    combined with conventional analytical properties of the wave
    function validate the RH.

    \section{Escape effect}
    \label{inside}
    
    It is convenient to generalize notation in Eq.(\ref{gen}),
    presenting the wave function of a particle in the form
    \begin{eqnarray}
      \label{inout}
      | \psi \rangle = |\mathrm{in} \rangle +
      \mathcal{R} \, |\mathrm{out} \rangle~.
    \end{eqnarray}
    The first term here describes the wave function that has a
    conventional, incoming behavior in the vicinity of the horizon.
    The second term is a wave that has an unexpected, outgoing
    behavior on the horizon. The arguments discussed above verified
    Eq.(\ref{inout}) for the outside region.
    
    Importantly, this equation remains valid for the inside region as
    well. The proof of this later claim given in
    \cite{kuchiev_1,kuchiev_2,kuchiev_3} goes along the following
    lines.  First one recovers the time-dependence of the wave
    function in Eq.(\ref{gen}) by simply multiplying it by a factor
    $\exp(-i\varepsilon t)$, i. e. writing $|\psi\rangle\equiv
    \psi(r)\exp(-i\varepsilon t)$. Then one uses the Kruskal variables
    Eq.(\ref{U}),(\ref{V}) that allow one to present this wave function on
    the horizon more conveniently, as
    \begin{eqnarray}
      \label{psiUV}
    |\psi\rangle = \exp[-i \varepsilon \ln(V^2)]+\mathcal{R}
    \exp[i\varepsilon \ln(U^2)]~.  
    \end{eqnarray}
    The crossing of the horizon between, for example, regions I and II
    corresponds to the change of sign of $U$, see Fig. \ref{one}.
    Obviously, this change does not affect the general structure of the
    wave function in Eq.(\ref{psiUV}) that is even in $U$ and $V$.
    This fact makes the methods of derivation of Eq.(\ref{inout}),
    which were used above for the outside region, applicable for
    the inside region as well.
    
    This discussion justifies the fact that the wave function always,
    including the inside region, has an admixture of the outgoing
    wave. For particles confined inside the horizon Eq.(\ref{inout})
    results in a new, unexpected and interesting phenomenon.
    Conventional arguments state that a particle that comes inside the
    horizon would stay inside forever because all classical
    trajectories for this particle eventually lead to the singularity
    at $r=0$. In quantum description these incoming trajectories
    correspond to the first term in the wave function in
    Eq.(\ref{inout}), which gives the incoming behavior in the
    vicinity of the horizon. However, the second term in the wave
    function gives the outgoing behavior on the horizon. In the
    classical description this term corresponds to those classical
    trajectories that lead from the singularity at $r=0$ into the
    outside world.  The presence of the two terms in the wave function
    means therefore that the events that describe the incoming particle
    are necessarily mixed in the wave function with the events that
    describe the outgoing particle, as was found in
    \cite{kuchiev_1,kuchiev_2,kuchiev_3}. Simply speaking, the
    particle confined inside the horizon has a chance to escape into
    the outside world. We call this the escape effect (EE).

    \subsection{Hawking radiation}

    \label{hawking}
    
    Conventional qualitative explanation for the Hawking effect refers
    to the virtual particle-antiparticle pairs that exist in the
    vicinity of the horizon due to quantum fluctuations. The strong
    gravitational field on the horizon is able to separate the pair,
    bringing one of its components inside the black hole, and allowing
    the other component to go outside and constitute the flux of
    outgoing radiation.
        
    The EE provides a different, appealing explanation of the
    radiation phenomenon. The radiation happens due to the fact that a
    particle locked inside the horizon can escape from the confinement
    into the outside world, creating the flux of outgoing radiation.
    The EE is entirely related to the second term in the right-hand
    side of Eq.(\ref{psiUV}).  Accordingly, the probability that a
    particle escapes into the outside region is governed by a factor
    $\mathcal{P}$,
    \begin{eqnarray}
      \label{P}
    \mathcal{P} \,\propto\,
    |\mathcal{R}|^2 \,= \,\exp\left(-\frac{\varepsilon}{T}\right)~, 
    \end{eqnarray}
    which looks similar to the probability of the Hawking radiation.
    However, to make this similarity complete, one has to presume that
    The distribution of particles inside the event horizon is governed by
    the same Hawking temperature. This condition is satisfied when a
    black hole is put inside the temperature bath, which has the
    Hawking temperature. This formulation of the problem was
    discussed by Hartle and Hawking \cite{hartle_hawking_1976}. The
    analysis in \cite{kuchiev_3} shows that for this particular case
    the flux of radiation that appears due to the EE reproduces the
    spectrum of a black body with the Hawking temperature.

    \subsection{Escape of particles and information transfer from the 
      inside region}
    \label{inform}
    
    Conventionally it is presumed that an outside observer can measure
    only few characteristics of the black hole, such as its mass, spin
    and charge. All other information related to the collapsing matter
    that created the black hole is supposed to be screened from the
    outside observer by the event horizon. Thus, presumably the
    collapse produces large information loss.
    
    However, the wave function Eq.(\ref{psiUV}) indicates that there
    is the EE, which provides a chance to retrieve the information
    from within the horizon back into the outside world.
    
    To be specific, consider a situation when the usual matter (made
    of electrons, protons etc) collapses producing a black hole.
    Conventionally it is supposed that for the outside observer this
    black hole would not look different from a black hole made from
    the antimatter (positrons, antiprotons etc). However, the
    discussion in Section \ref{hawking} indicates that the radiation
    of black holes takes place due to the EE.  If the black hole is
    made of matter, then there are only electrons inside, but there
    are no positrons \footnote{Assuming that during the collapse the
      temperature of the matter remained sufficiently low for
      production of the antimatter in large quantities}. In this case
    the outside observer would be able to see the flux of electrons,
    which escape from the inside region, while there would be no
    positrons in this spectrum.  From this fact the observer concludes
    that the black hole is made from conventional matter.  Similarly,
    the outside observer is able to detect other signals that
    correspond to other characteristics of the collapsed matter, thus
    retrieving the information hidden inside the horizon.
    
    At this point it is instructive to return back and compare the EE
    with the phenomenon of the Hawking radiation. There are some
    similarities.  In both cases there is a flux of radiation due to
    processes that take place on the horizon, in both cases the
    exponential function of the ratio of energy to the Hawking
    temperature is present. However, there are serious distinctions.
    The Hawking process is often explained via the pair production on
    the horizon. This usual and clearly looking physical explanation
    possesses, though, an intrinsic difficulty. The component of the
    pair that goes inside the horizon should possess the negative
    energy. This is the only way that allows the black holes to reduce
    its mass in the process. This negative energy of the ingoing
    particle equals $\varepsilon_\mathrm{in} =- \varepsilon$, where
    $\varepsilon>0$ is the positive energy of the outgoing particle.
    The point is that $\varepsilon_\mathrm{in}$ is supposed to be the
    total energy of the ingoing particle, the energy which is
    conserved, being equal therefore the energy that a particle would
    possess when located far outside of the black hole. This energy
    should definitely be positive. Admitting that it is negative, one
    makes an assumption, which introduces a difficulty into this
    physical picture of the Hawking radiation \footnote{This
      difficulty was a driving force that prompted the studies in
      \cite{kuchiev_1,kuchiev_2,kuchiev_3}}.
    
    The EE does not have this problem. The EE states that a particle
    is able to escape over the horizon. During this process its
    positive energy remains intact. Obviously, when a particle is
    left, the mass of the black hole becomes smaller. There is no need
    in this physical picture to introduce negative energies.
    
    Another important feature that distinguishes the EE from the
    Hawking radiation is the actual spectrum of a black hole. For the
    EE process it is {\em not} governed by the temperature. The
    temperature, as it appears in Eq.(\ref{P}), is a parameter that
    has only limited applicability, describing the probability of the
    escape of the particle from the inside region. But the flux of the
    outgoing particles depends also on the probability that particular
    particles exist inside the horizon. In other words, if there is
    some particular type of particles inside the horizon, then these
    particles can escape, giving a contribution to the radiation
    spectrum. If, however, there are no particles of this particular
    type inside, then this type of particles does not contribute to
    the radiation spectrum.  Accepting this ``limited'' point of view
    on the temperature, one should, probably, also modify the point of
    view on the entropy and the thermodynamic properties of black
    holes, but we do not elaborate on this argument here leaving it for
    further considerations.

    \section{conclusion}
    \label{conc}
    
    The central point of the presented discussion is the claim of
    Refs. \cite{kuchiev_1,kuchiev_2,kuchiev_3} that the wave function
    of any particle that approaches a black hole has an admixture which
    possesses unusual properties, describing the outgoing wave on the
    horizon. This important property originates from pure quantum
    reasons.  In the classical approximation all trajectories cross
    the horizon smoothly, leaving no opportunity for a particle to
    switch from the incoming to outgoing trajectory. Following the
    incoming trajectories all particles inevitably end up in the
    singularity at $r=0$. In contrast, on the quantum level the
    incoming and outgoing waves are mixed in the wave function on the
    horizon. Thus, it is impossible to describe the particle in terms
    of the wave that has only incoming component on the horizon.
    
    The existence of the outgoing wave on the horizon has important
    and unexpected implications. One of them is related to the
    scattering problem. Any particle approaching the black hole can
    bounce on the horizon back into the outside world. The
    corresponding effect, called the reflection on the horizon,
    drastically reduces the absorption cross section in the infrared
    region \cite{kuchiev_flambaum_04}.
    
    Another notable phenomenon takes place for the collapsed matter
    that is confined inside the event horizon. Classically such
    confinement is absolute, there is no way for a particle to return
    into the outside world. The quantum treatment shows that there is
    such a chance, a particle can escape.  This opportunity was called
    the escape effect. The probability that some particle gets away is
    governed by the exponential factor, which looks very conventional,
    being dependent on the ratio of the energy of the particle to the
    Hawking temperature. Due to this reason the flux of escaping
    particles resembles the spectrum of the Hawking radiation.  This
    similarity turns into identity when we consider a black hole that
    is placed inside the temperature bath.
    
    However, in the general case the spectrum of the escaped particles
    depends also on properties of the collapsed matter. As a result
    the flux of the escaped particles brings the information from the
    inside region into the outside world. This important fact may,
    probably, help resolve the information paradox. We did not attempt
    to prove in this work (or in the previous ones
    \cite{kuchiev_1,kuchiev_2,kuchiev_3,kuchiev_flambaum_04}) that
    {\em all} the information about the collapsing matter can be
    retrieved from under the horizon. However, the escape effect
    definitely allows some information to be recovered.  The
    implications of this fact may be far reaching, prompting, probably, a
    new look on thermodynamics properties of black holes. 
    
    This work was supported by the Australian Research Council.

\end{document}